\documentclass[referee]{raa}            % referee version: for submission

\usepackage{graphicx,times}             %for PS/EPS graphics inclusion, new
\usepackage{natbib}
\usepackage{amssymb,amsmath}
\bibpunct{(}{)}{;}{a}{}{,}

\usepackage[a4paper=true,dvipdfm=true,pagebackref=true]{hyperref}
\hypersetup{colorlinks = true, linkcolor = green, anchorcolor = red, citecolor = blue, filecolor = red, pagecolor = red, urlcolor = red}

\def\sunm{M_{\odot}}

\begin{document}

   \title{Physical properties variation across the green valley for local galaxies
%\footnotetext{$*$ Supported by the National Natural Science Foundation of China.}
}
%   \subtitle{I. Place Your Subtitle Here}

   \volnopage{Vol.0 (20xx) No.0, 000--000}      %%preserved for Editor. DOn't remove!
   \setcounter{page}{1}          %%starting page, preserved for Editor. DOn't remove!

   \author{Xue Ge
   	\inst{1,2,3}
   	\and Qiu-Sheng Gu
   	\inst{1,2,3}
   	\and Yong-Yun Chen
   	\inst{1,2,3}
	\and Nan Ding
	\inst{1,2,3}
   }

   \institute{School of Astronomy and Space Science, Nanjing University, Nanjing 210093, P.~R.~China; {\it qsgu@nju.edu.cn}\\
        \and
             Key Laboratory of Modern Astronomy and Astrophysics, Nanjing University, Nanjing 210093, P.~R.~China\\
        \and
             Collaborative Innovation Center of Modern Astronomy and Space Exploration, Nanjing 210093, P.~R.~China\\
\vs\no
   {\small Received~~20xx month day; accepted~~20xx~~month day}}

\abstract{ We have selected a sample of nearby galaxies from Sloan Digital Sky Survey Data Release 7 (SDSS DR7) to investigate the physical properties variation from blue cloud to green valley to red sequence. The sample is limited in a narrow range in color-stellar mass diagram. After splitting green valley galaxies into two parts---a bluer green valley (green 1) and a redder one (green 2) and
three stellar mass bins, we investigate the physical properties variation across the green valley region. Our main results are as following:
(i) The percentages of pure bulge and bulge-dominated/elliptical galaxies increase gradually from blue cloud to red sequence while the percentages of pure disk and disk-dominated/spiral galaxies decrease gradually in all stellar mass bins and different environments;
(ii) With the analysis of morphological and structural parameters (e.g., concentration (C) and the stellar mass surface density within the central 1Kpc ($\Sigma_{1}$)), red galaxies show the most luminous and compact cores than both green valley and blue galaxies while blue galaxies show the opposite behavior in all stellar mass bins.
(iii) A strong negative (positive) relationship between bulge-to-total light ratio (B/T) and specific star formation rate (sSFR) ($D_{4000}$) is found from blue to red galaxies.
Our results indicate that the growth of bulge plays an important role when the galaxies change from the blue cloud, to green valley, and to the red sequence.
}

   \authorrunning{Xue Ge et al. }            %author_head in even pages
   \titlerunning{Properties variation across the green valley }  % title_head in odd pages

   \maketitle
%% The author head (on even pages) and the title head (on odd pages) will be
%% automatically extracted from \author{} and \title{}. Whenever the title is too long,
%% you will be asked to supply a shorter one by inserting either \authorrunning{} or
%% \titlerunning{} before \maketitle. Anyway, you can specify your own heads.
%%
%%
%% Note: In the following text body of your manuscript, please note several differences from
%%       other major journals:
%% (1) \subsection{Please Capitalize the First Letter of Each Notional Word in Subsection Title}
%% (2) Please Capitalize the First Letter of Each Notional Word in all tables' captions

%
%________________________________________________ sections below
%
\section{Introduction}           %% first-level sections will be auto-capitalized
\label{sect:intro}

A bimodal distribution of optical color \citep{Strateva2001, Blanton2003}, ultraviolet-optical color \citep{Salim2007}, morphologies \citep{Driver2006} or star formation rates (SFR) \citep{Kauffmann2003a, Kauffmann2003b} of galaxies has been unambiguously observed. In color-magnitude diagrams (CMD), the galaxies are divided into ``red sequence'' and ``blue cloud''. Generally speaking, red sequence contains old and quiescent galaxies \citep{Kauffmann2003a}, while blue cloud mainly contains blue star-forming disky galaxies \citep{Kaviraj2014a, Kaviraj2014b}. The galaxies between red sequence and blue cloud are called ``green valley'', which are considered as a transition population. \cite{Wyder2007} and \cite{Jin2014} found that the two-Gaussian fitting to the galaxies in CMD is not sufficient which suggests that the green valley is not a simple mixture of red sequence and blue cloud. Therefore, green valley galaxies can provide us with crucial clues to connect red sequence and blue cloud in terms of star formation quenching and evolution of galaxies.

Previous studies have shown that since $z\sim1$ the number and stellar mass of blue galaxies are almost constant while the stellar mass of red galaxies have increased by a factor of 2 $\sim$ 4 \citep{Bell2004, Faber2007}. This scenario supports that the existence of red galaxies requires certain quenching mechanisms to stop or weaken the star formation in blue galaxies \citep{Bell2004}. Different quenching mechanisms have been proposed to explain the transition from blue to red galaxies,
such as major mergers \citep{Springel2005,Di Matteo2005}, AGN and supernovae feedback \citep{Di Matteo2005, Nandra2007, Marasco2012}, morphological quenching \citep{Martig2009, Martig2013} and environment quenching \citep{Peng2010}.

The quenching mechanisms mentioned above propose restrictions on the galaxy structure. These restrictions provide us with an additional approach to understand the connection between the changes of galaxy structure and the locations that galaxy resides in. For example, elliptical galaxies are more concentrated due to internal or external processes than spiral galaxies. This leaves a motivation for us to explore the connection between the morphology/structure and the star formation activity. \cite{Pan2013} have shown that green valley galaxies have the lower (higher) Gini coefficient (G) (second order moment (M20)) than red galaxies but higher (lower) G (M20) than blue galaxies. The average value of asymmetry parameter (A) for green valley galaxies is also between both red and blue galaxies. Moreover, the strong connection between morphological/structural parameters and star formation activity has been demonstrated in the studies of local galaxy surveys \citep{Kauffmann2003b, Mendez2011, Fang2013}. Some structural thresholds, such as critical stellar mass, stellar surface mass density and S\'{e}rsic index reflect the transformation from blue galaxies to red galaxies. Galaxies above these thresholds tend to be old or quiescent while galaxies tend to be young or active below these thresholds \citep{Kauffmann2003b, Driver2006, Schiminovich2007, Bell2008}. \cite{Cheung2012} and \cite{Fang2013} used a structural parameter, $\Sigma_{1}$ (the stellar mass surface density within the central 1kpc), to investigate whether there is a difference between different galaxy colors, and found $\Sigma_{1}$ is a better indicator for the sequence of galaxies than other parameters. \cite{Bait2017} used multiwavelength data to study the dependence of star formation on the morphology and environment in local universe. Their results suggested that the morphology of galaxies correlates with star formation although environmental effects on morphology are weak.

This work will focus on the physical properties variation from blue cloud to green valley to red sequence in a narrow range of optical color and different stellar mass bins. In particular, we try to investigate whether there is a monotonic variation in morphology/structure and star formation, so we split green valley galaxies into two populations (i.e., green 1 and green 2). We also expect to have a further understanding to the environmental effects on star formation activity in the local universe. The outline of the paper is as follows. In section 2, we introduce our sample selection and data. The results and discussions are presented in section 3 and 4. Finally, we present our conclusions in section 5. The cosmological constants adopted in this work are $\Omega_m=0.3, \Omega_\Lambda=0.7$ and $H_0=70$ $km^{-1}$ $s^{-1}$ $Mpc^{-1}$.

%% Authors can give a citation as 'Michel et al. 1992'.
%% You may also use \cite, \citep and \citet for citation, and use Table~1 or Figure~1
%% and so forth. Using \ref and \label for cross-references of Tables/Figures
%% is a good way in adjusting/adding/removing text, tables or figures.

\section{Sample and Data}
\label{sect:Samm}

\begin{figure}
\centering
\includegraphics[width=0.8\textwidth]{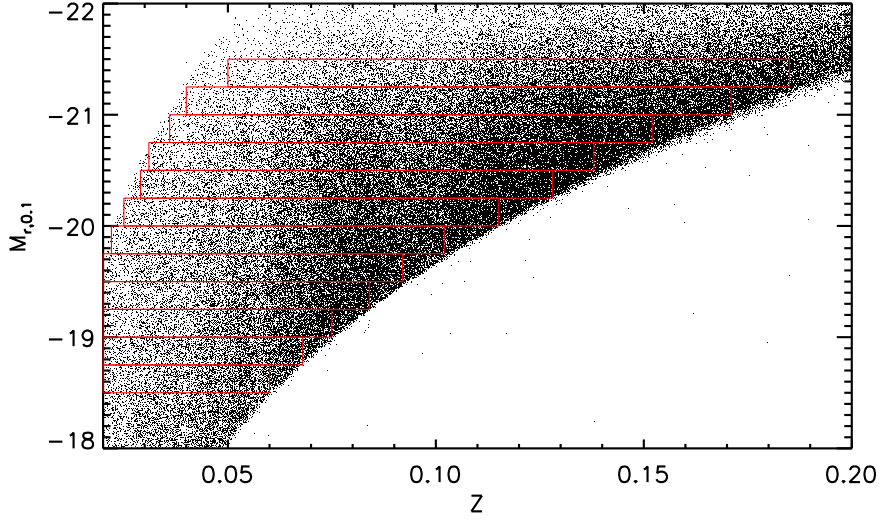}
%\vspace{-4em}
\caption{The preliminary sample distribution (black points) in r band absolute magnitude and $z$ space. $M_{r,0.1}$ is the absolute magnitude with k-corrected to $z = 0.1$ in r band. We select objects only in the red boxes to produce the final sample analyzed in this paper.}
\label{f1}
\end{figure}

\cite{Meert2015} and \cite{Meert2016} presented a catalog containing 670722 galaxies selected from Sloan Digital Sky Survey Data Release 7 (SDSS DR7) \citep{York2000, Abazajian2009}. These galaxies were chosen based on three criteria:  (1) 14 $<$ r band Petrosian magnitude $<$ 17.7 after the Galaxy extinction correction ; (2) the object should be a galaxy with the identification of the photo pipeline; (3) the spectrum of the object is recognized as a galaxy. Each galaxy was fitted by using four models (de Vacouleurs (Dev), S\'{e}rsic (Ser), de Vacouleurs+Exponential (DevExp), and S\'{e}rsic+Exponential (SerExp) profiles) with the point spread function-corrected in g, r and i band respectively. The routine of Galfit \citep{Peng2002} and analysis pipeline PyMorph \citep{Vikram2010} were used for the fitting.
We choose the SerExp model where a galaxy was fitted with two components (bulge and disk). We select the galaxies with $n < 8$ (the S\'{e}rsic index of bulge), $b/a > 0.63$ (the ratio of minor axis to major axis) in g, r, and i band and flag bits 1 - 13 in r band \citep{Meert2015,Meert2016}. In order to acquire k-corrected Petrosian magnitudes, we cross match the catalog with the New York University Value-Added Catalog \citep{Blanton2005} and derive preliminary sample containing 155388 galaxies. Figure \ref{f1} shows the preliminary sample selected according to the criteria above. To ensure a broader span of luminosity and the completeness of color in each magnitude bin, we divide galaxies into 12 bins from $M_{r,0.1} = -18.50 \sim -21.5$ mag and $0.02 < z < 0.18$ with bin size of 0.25 mag, as shown in Figure \ref{f1} ($M_{r,0.1}$ is the absolute magnitude in r band k-corrected to $z$ = 0.1).
The sources with quite high or low r band absolute magnitudes are excluded and we use only the sources in the red boxes to produce the final sample.

\begin{figure}
\centering
%\hspace{-4ex}
\includegraphics[width=0.8\textwidth]{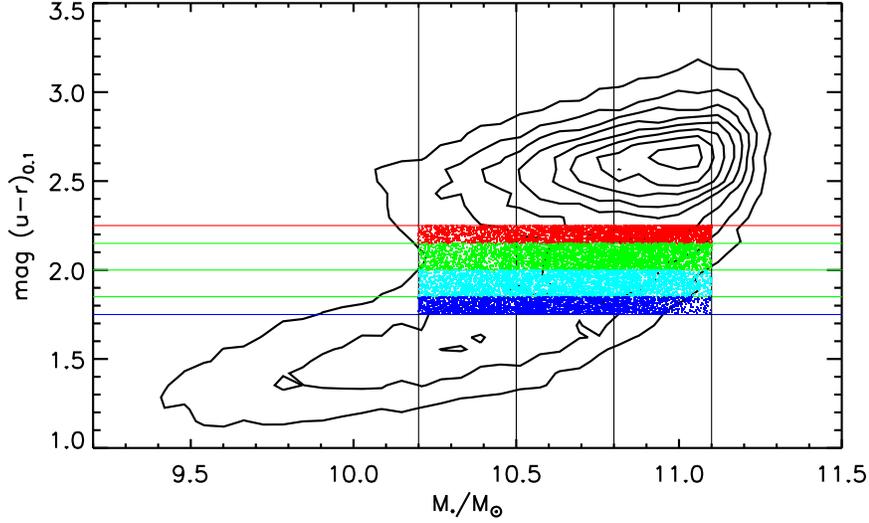}
\caption{The distribution of color-stellar mass for galaxies in red boxes in Figure \ref{f1}. $(u-r)_{0.1}$ is the color of k-corrected to $z = 0.1$. The galaxies with $(u-r)_{0.1}$ color between 1.75 and 1.85 (blue points), 1.85 and 2.0 (cyan points), 2.0 and 2.15 (green points) and 2.15 and 2.25 (red points) are defined as blue galaxies, green 1 galaxies, green 2 galaxies and red galaxies respectively. The vertical lines represent the different stellar mass bins.}
\label{f2}
\end{figure}

Figure \ref{f2} shows the distribution of color-stellar mass for galaxies selected in Figure \ref{f1}. $(u-r)_{0.1}$ is the color of k-corrected to $z = 0.1$ and stellar masses are obtained from MPA/JHU \footnote{http://www.mpa-garching.mpg.de/SDSS/DR7/}\citep{Kauffmann2003a}. We select the sources with $(u-r)_{0.1}$ color from 1.75 to 2.25 and stellar mass from 10.2 to 11.1 (in logarithm scale) as our final sample. We define the galaxies with $(u-r)_{0.1}$ color between 1.75 and 1.85, between 1.85 and 2.0, between 2.0 and 2.15 and between 2.15 and 2.25 as blue galaxies, green 1 galaxies, green 2 galaxies and red galaxies respectively. Further, the stellar mass is spelt into three ranges (i.e., [10.2,10.5], [10.5,10.8], [10.8,11.1]) to investigate the role stellar mass plays when the blue galaxies go through green valley into red sequence. Table \ref{tab1} provides the information of the final sample.
Considering the contamination of dusty star-forming galaxies to green valley galaxies, we estimate the fraction of dusty green valley galaxies in our sample based on WISE Mid-infrared \citep{Cutri2014} and GALEX NUV
\citep{Martin2005} catalogs. The dusty star-forming galaxies are defined as galaxies with $f_{12\mu m}/f_{nuv} > 200$ and $f_{4.6\mu m}/f_{3.4\mu m} > 0.85$ \citep{Yesul2014}. We find there are only 2 percent dusty star-forming galaxies in green valley, which is negligible and will not affect our results significantly.
%\textcolor[rgb]{1.00,0.00,0.00}

Physical parameters include: morphological parameters (concentration (C), M20 and G) and structural parameters (n and bulge-to-total light ratio (B/T)) obtained from \cite{Meert2015}, Specific star formation rate (sSFR) and $D_{4000}$ obtained from MPA/JHU catalog.

\begin{table}[!h]
\centering\caption{The galaxies in our final sample.
Col. (1): Galaxy type;
Col. (2): Stellar mass range;
Col. (3): $(u-r)_{0.1}$ color with k-corrected to $z = 0.1$;
Col. (4): Galaxy number.}
\label{tab1}
\begin{tabular}{cccc}
\hline\hline
Galaxy type & $log M_*$ & $(u-r)_{0.1}$ & Number\\
\hline
            &  [10.2 - 10.5] & [1.75 - 1.85] & 1110 \\

Blue     &  [10.5 - 10.8] & [1.75 - 1.85] & 1263 \\

            &  [10.8 - 11.1] & [1.75 - 1.85] & 692 \\
\hline

            & [10.2 - 10.5] & [ 1.85 - 2.0 ]  & 1273 \\

Green 1 & [10.5 - 10.8] & [ 1.85 - 2.0 ]  & 1668 \\

            & [10.8 - 11.1] & [ 1.85 - 2.0 ]  & 1328 \\

\hline
            & [10.2 - 10.5] & [ 2.0 - 2.15 ]  & 938 \\

Green 2 & [10.5 - 10.8] & [ 2.0 - 2.15 ]  & 1625 \\

             & [10.8 - 11.1] & [ 2.0 - 2.15 ]  & 1682 \\
\hline

            & [10.2 - 10.5] & [2.15 - 2.25] & 689 \\

Red      & [10.5 - 10.8] & [2.15 - 2.25] & 1160 \\

            & [10.8 - 11.1] & [2.15 - 2.25] & 1491 \\

\hline\hline
\end{tabular}
\end{table}

\section{Result}
\label{sect:result}
\subsection{Morphological distribution}

\begin{figure}
\centering
\includegraphics[width=1.0\textwidth]{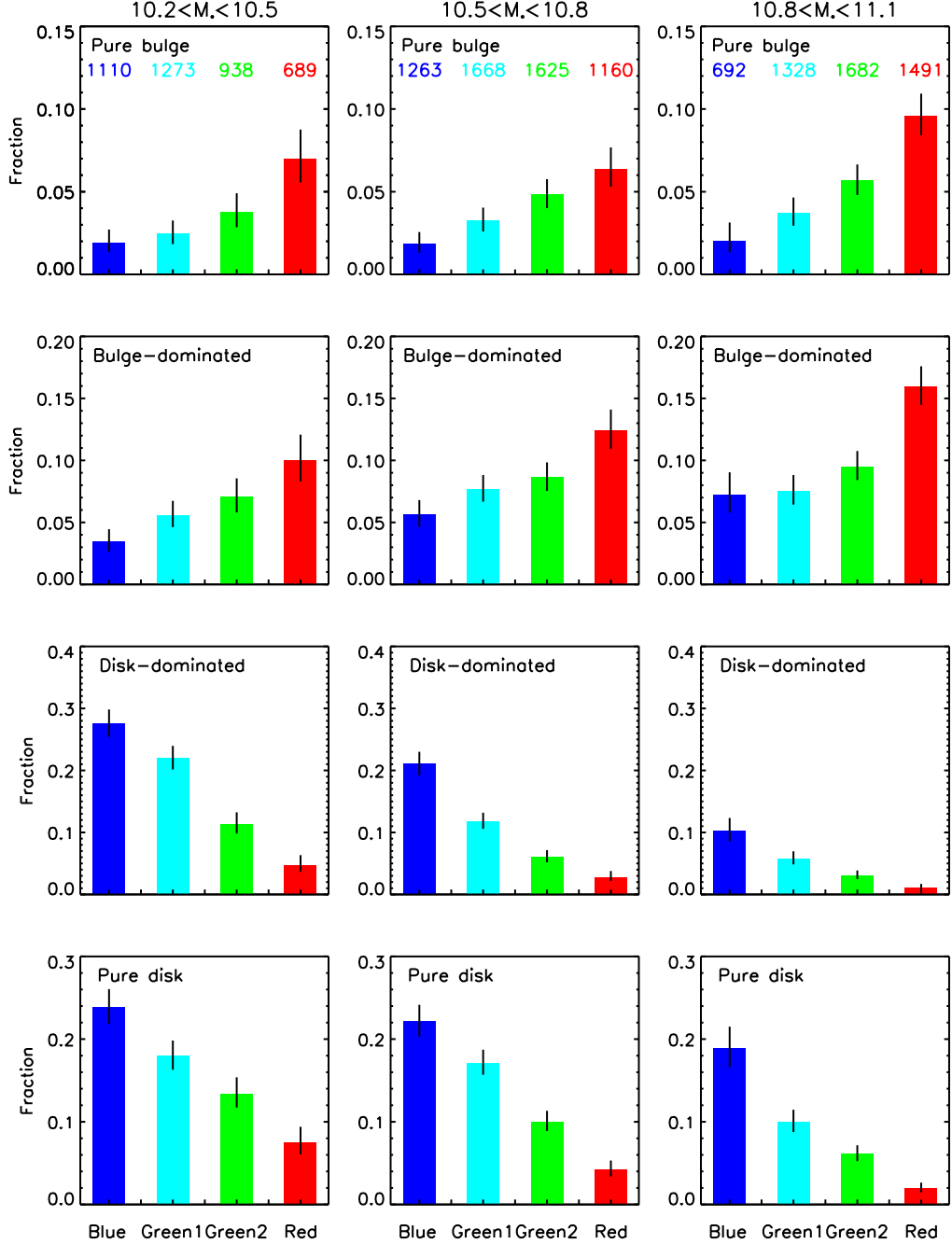}
\vspace{-2em}
\caption{The fractions of different morphologies (from top to bottom: pure bulge, bulge-dominated, disk-dominated and pure disk) for blue, green 1, green 2 and red galaxies in $10.2 < log M_* < 10.5$ (left), $10.5 < log M_* < 10.8$ (middle), $10.8 < log M_* < 11.1$ (right). The identification of different morphologies is from the bulge and disk decomposition. The black vertical lines are the binomial error of fraction \citep{Cameron2011}. The number of galaxy types in each stellar mass bin is marked on the top (see also Table \ref{tab1}). }
\label{f3}
\end{figure}

We separate galaxies into five morphologies based on the flagged system in \cite{Meert2015}. In the flagged system, bulge galaxies and disk galaxies have flag bits 1 and 4 set respectively. Two-component galaxies have flag bits 10 set. For bulge galaxies, we define galaxies with flag bits 2 set as pure bulge while flag bits 3 set as bulge-dominated. For disk galaxies, pure disk and disk-dominated galaxies have flag bits 5 set and 6, 7, 8, 9 set respectively. The remaining galaxies are defined as two-component galaxies.
Figure \ref{f3} shows the distributions of different morphologies in different stellar mass bins. We can see clearly the fractions of pure bulge and bulge-dominated galaxies increase gradually while the fractions of pure disk and disk-dominated galaxies decrease gradually from blue to green to red galaxies in all stellar mass bins. What is more, it is found that the percentage of bulge galaxies appears to be increasing tendency from low to high stellar mass bins which is independence of the galaxy colors. In contrast, the percentage of disk galaxies shows a decreasing tendency as we move from low to high stellar mass ranges. Obviously, these variations will lead to the buildup of massive bulge galaxies and the insufficiency of massive disk galaxies.

Many pervious studies involving the multi-component decomposition of large sample galaxies \citep{Simard2011, Lackner2012, Meert2015,Meert2016} are model-dependent. So, we cross match our sample with the catalog of Galaxy Zoo 1 \citep{Lintott2011} which listed the morphological classifications of all galaxies with $0.001 < z < 0.25$ in SDSS DR7. Six morphologies (Elliptical galaxy, Clockwise/Z-wise spiral galaxy, Anti-clockwise/S-wise spiral galaxy, Spiral galaxy other, Star or Unknown and Merger) based on visual classifications were derived and each object was classified as a possible morphology 38 times on average by thousands of volunteers. We combine Clockwise/Z-wise spiral galaxy, Anti-clockwise/S-wise spiral galaxy and Spiral galaxy other as spiral classification. Moreover, \cite{Yang2007} provided  a catalog of environment where the masses of dark matter halo for SDSS DR4 galaxies have been estimated. We follow the way provided by \cite{van den Bosch2002} and define the environment of galaxies based on dark matter halo masses. A galaxy is defined as field galaxy, group galaxy or cluster galaxy if its dark matter halo mass is less than $10^{13}\sunm$, between $10^{13}\sunm$ and $10^{14}\sunm$ or higher than $10^{14}\sunm$, respectively.
We do not choose a threshold to classify a galaxy as a elliptical or spiral considering a conservative threshold can reduce significantly the number of sample. Actually, the medians of votes that have been corrected for classification bias \citep{Bamford2009} can also reflect the morphologies of galaxies.

\begin{figure}
\centering
\includegraphics[width=1.0\textwidth]{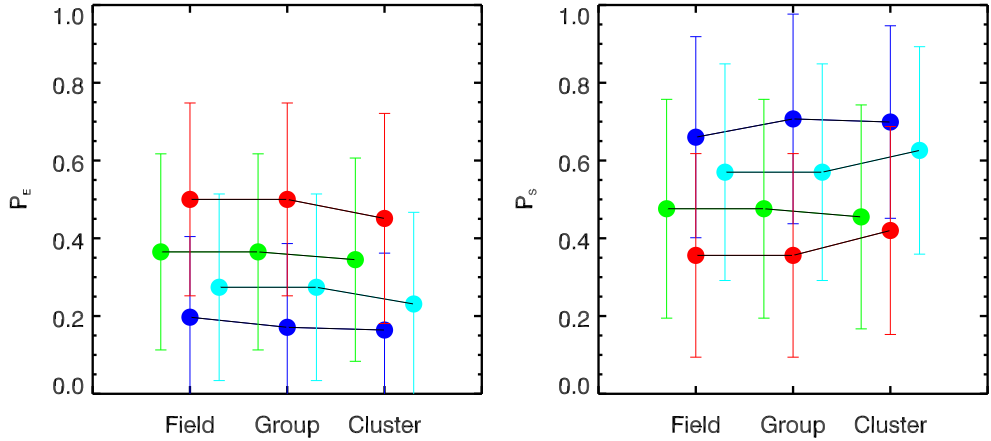}
\vspace{-6em}
\caption{Median probabilities of elliptical (left) and spiral (right) vs different environments. Blue galaxies, green1 galaxies, green 2 galaxies and red galaxies are marked with blue, cyan, green and red solid circles respectively. At each panel, the standard deviations are given as the error of median probabilities. Green and cyan points are shifted in 0.03 along x-axis.}
\label{f4}
\end{figure}

Figure \ref{f4} displays the relationship between the medians of votes and environments for different galaxy sequences. We conclude that blue galaxies have the higher probability to be classified as spiral galaxies while the red galaxies have the higher probability to be classified as elliptical galaxies. The median probabilities of being spiral or elliptical for green 1 and green 2 galaxies are between blue and red galaxies. This result is consistent with Figure \ref{f3}. It notes that we have displayed that the variation of morphological distribution is independent of stellar mass (Figure \ref{f3}). So in Figure \ref{f4}, we do not split galaxies into different stellar mass bins. We do not find the environment can strongly change the probability of being a morphological classification. For different environments, it is found that the fraction of green valley galaxies is approximately constant at 57-61 percent although we select only a part of blue and red galaxies. In addition, the ratio of blue to green valley galaxies decreases significantly with environmental richness (from 38 percent in field to 20 percent in cluster). These results support the criterion where while environment does not affect significantly the morphological distribution of galaxies and the timescale of crossing green valley, it can urge the process by which galaxies start evolve from blue cloud to red sequence \citep{Bremer2018}.

It is well accepted that blue galaxies are active star-forming galaxies and the most of them are dominated by disks. On the other side, red galaxies are quiescent elliptical galaxies with compact cores. Green valley galaxies have morphological properties intermediate between blue and red galaxies. The discoveries above is consistent with \cite{Coenda2018} and the variation of morphology is monotonic from blue to green 1 to green 2 to red galaxies.
%support the conclusion that  (\cite{Balogh2011, Mendez2011, Pan2013}).

\subsection{Morphological and structural parameters}

\begin{figure}
\centering
\includegraphics[width=0.9\textwidth]{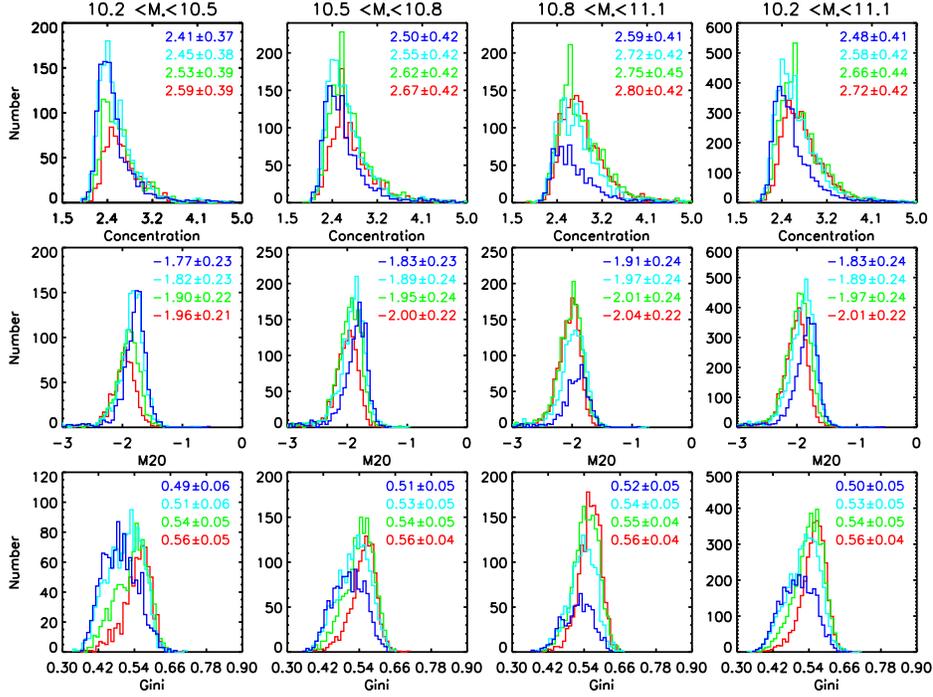}
\vspace{-1em}
\caption{The distributions of different morphological parameters including C, M20 and G (from top to bottom) in $10.2 < log M_* < 10.5$, $10.5 < log M_* < 10.8$, $10.8 < log M_* < 11.1$ and $10.2 < log M_* < 11.1$ (from left to right). Blue, cyan, green and red histograms represent blue, green 1, green 2 and red galaxies respectively. The corresponding medians and standard deviations are labeled on the top right corner.}
\label{f5}
\end{figure}

Many previous work claimed that early type galaxies (ETGs) and late type galaxies (LTGs) have different distribution in G
and A diagrams \citep{Abraham1996, Lotz2004, Kong2009}. These results argued that ETGs have higher G and lower A than LTGs. In order to further confirm the tendency of variation for these parameters, we investigate the variation of the morphological parameters including C, M20 and G (Figure \ref{f5}). C parameter quantifies the concentration of star light in a galaxy. M20 and G represent the second order moment of the brightness 20 percent pixels and the light distribution of pixels in a galaxy (see the detail definition in \cite{Lotz2004}). We find a graded variation tendency from blue to green 1 to green 2 to red galaxies in all stellar mass bins. For example, the median of C changes from 2.41 to 2.59,  from 2.50 to 2.67, from 2.59 to 2.80 in three stellar mass bins respectively. In entire blue, green valley and red galaxies, the median is from 2.48 to 2.72. This tendency of variation is found not only in C but also in M20 and G. The K-S test shows these distributions have significant difference with p $\ll$ 0.0001. We can conclude that the variation of these morphological parameters results from the prominence of galaxy nuclei.

NASA-Sloan Atlas \footnote{http://www.sdss.org/dr14/manga/manga-target-selection/nsa/} provides the surface brightness profiles of SDSS local galaxies in u, g, r, i, z bands in series of angular size. In this work, we compute stellar mass surface density profiles, $\Sigma_{1}$, utilizing this catalog. The computational processes are listed as following steps. Firstly, the cumulative light profile in i band is fitted through spline algorithms which allow us to obtain the interpolated light profile at 1 Kpc. Secondly, the light within the central 1 Kpc is corrected by using SDSS Galactic extinction and k-corrected calculator of \cite{Chilingarian2012}. Finally, the surface brightness within the central 1 Kpc is converted into the mass density through the mass-to-light ratio (M/L) in i band. The stellar masses in M/L are obtained from MPA/JHU catalog \citep{Kauffmann2003a}. Population properties, such as metallicity and initial mass function, may affect the accuracy of M/L. However, \cite{Fang2013} have suggested that metallicity has nearly no effects on the measurement of M/L. The initial mass function would lead to the underestimate of $\Sigma_{1}$ of massive red galaxies, which would be in line with our expectations.

\begin{figure}
\centering
\includegraphics[width=0.9\textwidth]{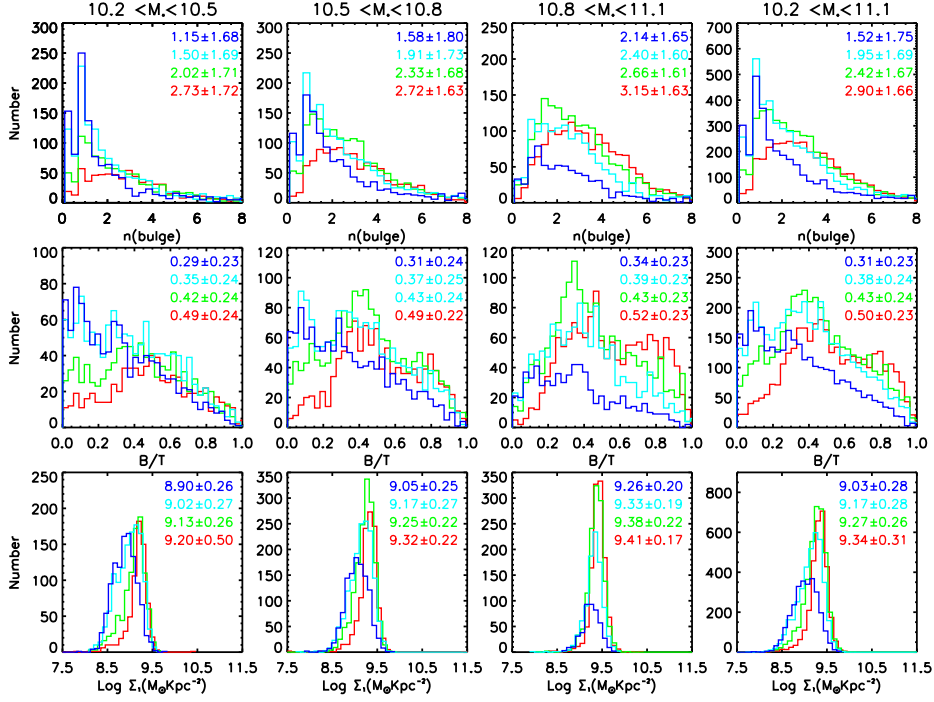}
\vspace{-2em}
\caption{The distributions of structural parameters including n, B/T and $\Sigma_{1}$ (from top to bottom). All colors and labels are the same as Figure \ref{f5}.}
\label{f6}
\end{figure}

In addition to $\Sigma_{1}$, we investigate also the variation of other structural parameters including n and B/T. Figure \ref{f6} shows the distributions of these parameters in different stellar mass bins. We find blue galaxies have the lowest medians of these parameters while red galaxies have the highest value and the phenomenon is independent of stellar mass. The K-S test with p $\ll$ 0.0001 shows these parameters have different distribution.
The continuous variation tendency from blue to green 1 to green 2 to red galaxies reflects a criterion in which the bulge component is more and more distinct. It indicates that bulge plays a important role during the transformation from blue cloud to red sequence.

\subsection{B/T and sSFR}
\begin{figure}
\centering
\includegraphics[width=0.8\textwidth]{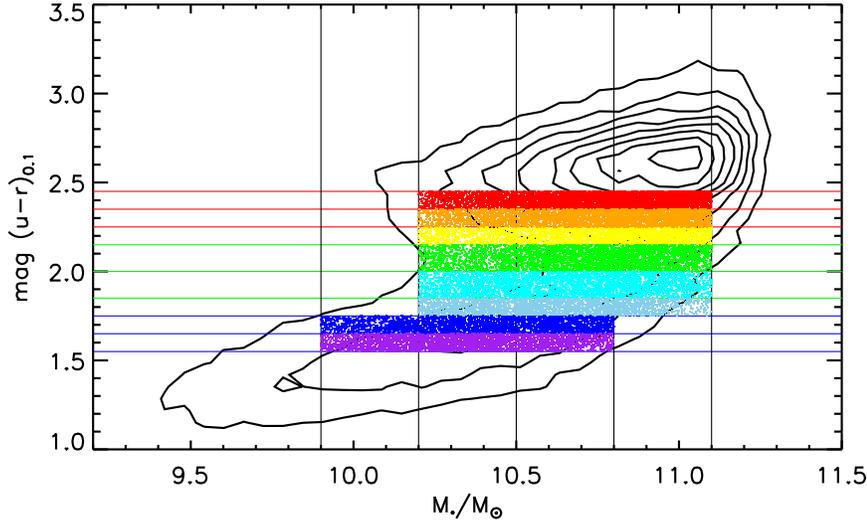}
\caption{Reconstructed sample containing three groups of blue galaxies (purple, blue and sky blue), two groups of green valley galaxies (cyan and green) and three groups of red galaxies (yellow, orange and red) in color-stellar mass diagram.}
\label{f7}
\end{figure}

\begin{figure}
\centering
\includegraphics[width=0.8\textwidth]{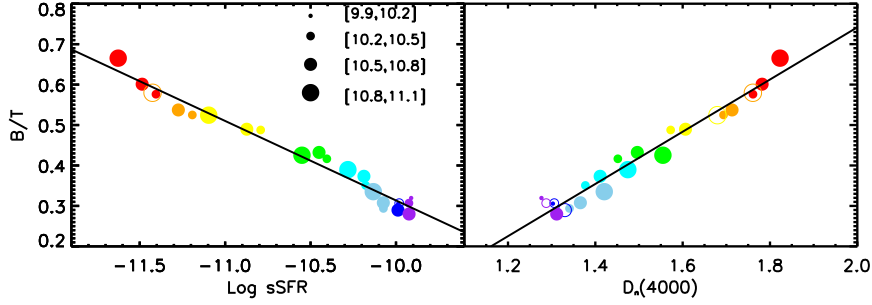}
\caption{The relationships between B/T and sSFR (left) / $D_{4000}$ (right). The size and color of circles represent the stellar mass and the range of $u-r$ respectively. Some points are marked with open circles because of the overlapping between the solid points. The solid black line in each panel represents the best fitting.}
\label{f8}
\end{figure}

In order to investigate the relationship between bulge and star formation activity from blue to red galaxies, we select two groups of additional blue galaxies and two groups of additional red galaxies. Figure \ref{f7} displays the reconstructed sample where purple, blue, yellow and red points are the additional sources used next. It notes that the stellar mass range for purple and blue points is from 9.9 to 10.8 in logarithm scale because of the lack of massive blue galaxies. In Figure \ref{f8} (left), we show the relationship between median B/T and sSFR. The solid black line in the panel is the best fitting. The formula of best fitting is
\begin{eqnarray}
B/T=-1.65(\pm0.08)-0.20(\pm0.01)  Log sSFR
\label{eq2}
\end{eqnarray}
We can see from the picture that B/T increases as the decreasing of sSFR from blue to red galaxies. Moreover, massive galaxies have the lower sSFR and higher B/T in the same $u-r$ color in general. We find that the extremely blue galaxies (blue and purple) do not follow the tendency. We contribute the deviation to the scatter of bulge-disk decomposition due to the inapparent bulges in extremely blue galaxies. sSFR used in this paper is computed by different methods \citep{Brinchmann2004}, which can generate some biases. As a substitute for sSFR, we show the relationship between median B/T and $D_{4000}$ (Figure \ref{f8} (right)). $D_{4000}$ is defined as the ratio of continuum 3850-3950 \AA\ and 4000-4100 \AA\ \citep{Balogh1999} and it is a good indicator for galaxy age, so it reflects the recent star formation activity in galaxies. A strong positive correlation is found between them. The formula of best fitting is
\begin{eqnarray}
B/T=-0.55(\pm0.05)+0.64(\pm0.03)  D_{4000}
\label{eq2}
\end{eqnarray}
sSFR or $D_{4000}$ can be considered as a proxy for the recent star formation. Lower sSFR or higher $D_{4000}$ suggests older stellar population. As we can see from Figure \ref{f8}, the galaxy with redder color and higher stellar mass have lower sSFR, which coincides with the buildup of bulge component. Our result is consistent with \cite{Salim2009} who found that sSFR have a continuous and smooth change in sSFR and rest-frame UV-optical color space in local galaxies.

\section{Discussion}
\label{sect:discussion}

Various studies have suggested that the morphology of galaxies (e.g., prominent bulge) is correlative with star formation history.
Simulations have shown that major mergers can lead to a starburst in the centre of galaxies through sweeping gas into centre \citep{Di Matteo2005}. The starburst consumes gas rapidly, which leads the growth of bulge and changes the morphology ultimately \citep{Springel2005, Hopkins2010, Cheung2012}. Red galaxies have higher proportion of bulge/elliptical galaxies in our study, which suggests that the prominent bulge in red galaxies may play an important role during the process of preventing star formation. For example, the existence of stellar spheroid is responsible for the stability of disks, thus against gravitational instability to form stars \citep{Martig2009, Martig2013}.
\cite{Cheung2012} and \cite{Fang2013} investigated the relationship between quenching mechanisms and galaxy structural parameters. They found that red galaxies have the highest $\Sigma_{1}$ and blue galaxies have the lowest value on average. Green valley galaxies are between blue and red galaxies. The suggestion they proposed was that $\Sigma_{1}$ may be more physically corrective with quenching process. However, the threshold is only a necessary condition in quenching process. It remains a ambiguous question for the connection between bulge growth and star formation quenching.
Combining $low-z$ with $high-z$ galaxies drawn from GAMA and CANDELS respectively, \cite{Brennan2017} studied the connection between the location relative to star formation main sequence (SFMS) and structural parameters. Based on a semi-analytic model in which the formation of bulge occurred not only due do merger events but also disk instabilities, they found that the star formation activity was strongly associated with the buildup of bulge component.

Our analysis based on the decomposition of bulge and disk and morphological/structural parameters show that the bulge component is more and more prominent when we go from blue to green 1 to green 2 to red galaxies. The monotonic variation is reflected through the largest $\Sigma_{1}$, n and B/T in red galaxies and smallest ones in blue galaxies. In addition, when we expand the sample, strong correlations between B/T and sSFR/$D_{4000}$ show that the recession of star formation is strongly correlative with the buildup of bulge. Although not all red galaxies have prominent bulge component, we can conclude that the recession of star formation is accompanied with the buildup of bulge. Just like \cite{Bell2012} and \cite{Lang2014}, they argued that the prominent bulge is a better indicator for the quenching of star formation.  Our results in this work are consistent with \cite{Brennan2017} who found also the monotonic variation tendency of galaxy structural parameters as the changes of location relative to SFMS.

We combine the morphological classification of Galaxy Zoo \citep{Lintott2011} with the environment catalog \citep{Yang2007} and investigate the effects of environment on the morphologies. Our result shows that the morphology is almost independent of the environment, which is consistent with \cite{Coenda2018}. Furthermore, the stellar mass of our sample is $> 10^{10} \sunm$. It is not strange that the quenching mode is more dependent on the internal processes rather than external processes \citep{Peng2010}. We find the ratio between green valley galaxies and total galaxies is approximately constant of 58 percent in different environments although we do not include all blue and red galaxies in our sample. It notes that the result is consistent with \cite{Bremer2018} who used $u-r$ color and stellar mass to define green valley galaxies. They found the ratio of green valley galaxies to the whole sample is approximately 18 percent at a given environmental richness. \cite{Bait2017} found that the fraction of green valley galaxies is almost constant (20 percent) in all environments based on the selection of sSFR. The invariable fraction indicates that the environment can not change the timescale for crossing green valley. Many previous studies have found the transition timescale is $\sim$1 Gyr otherwise we can not see the bimodal distribution in CMD diagram \citep{Faber2007, Balogh2011}.

It has been known that the stellar mass is strongly related to SFR in local and $high-z$ universe \citep{Brinchmann2004, Salim2007, Elbaz2007, Peng2010}.  In this paper, the galaxies are divided into three stellar mass bins (i.e., $10.2 < log M_* < 10.5$, $10.5 < log M_* < 10.8$, $10.8 < log M_* < 11.1$). We find the fractions of pure bulge and bulge-dominated galaxies are slightly higher in larger stellar mass than in smaller stellar mass. Simultaneously, pure disk and disk-dominated galaxies reside in more easily low stellar mass (see Figure \ref{f3}). The reduction of blue pure disk and disk-dominated galaxies coincides with the buildup of red pure bulge and bulge-dominated galaxies. It provides us with insight into the evolution of galaxy from blue to red. \cite{Faber2007} proposed a ``mixed'' model for the formation of spheroidal galaxies. In the model, the quenching progress can occur very early or late. In short, blue star-formation galaxies can quench to red massive galaxies through two ways. The first is very early quenching at low stellar mass such that fewer galaxies remain blue when reaching high stellar mass region. The second is very late quenching which occurs shortly after reaching enough high stellar mass.

In Figure \ref{f5} and Figure \ref{f6}, the medians of morphological and structural parameters have a mitigatory rise or decline from low to high stellar mass. It indicates that for galaxies with the same color, massive galaxies have more prominent bulge component (see also Figure \ref{f8}). Red galaxies have the more prominent bulge properties in higher stellar mass region. It seems that galaxies within high stellar mass range quench preferentially which will lead to the increasing of the number of massive red galaxies. Furthermore, less massive star formation galaxies will shut down their star formation in later epoch and shift to the end of massive quenching galaxies. It is consistent with the criterion of ``Downsizing'' \citep{Cowie1996}.

\section{Conclusions}
\label{sect:conclusion}

Using a sample drawn from SDSS DR7, together with the data from other public catalogs, we have investigated the physical parameters variation for blue, green 1, green  2 and red galaxies. In addition, we have also discussed the importance of bulge in quenching star formation. Our main conclusions are as follows:

(1) The blue galaxies have the lowest fraction of bulge/elliptical galaxies and the highest fraction of disk/spiral galaxies. However, the fraction is opposite in red galaxies. Green valley galaxies have intermediate fraction. The morphologies show continuous change from blue cloud to green valley to red sequence. It is found that the monotonic variation is independent of the stellar mass and environments.

(2) The fraction of green valley galaxies is almost constant while the ratio between blue and green valley galaxies decrease as we move from low to high environmental richness, which indicates that the effects of environment is not on the timescale of crossing green valley but on the time when the blue galaxies start to transform their morphologies.

(3) The medians of morphological parameters (C, M20 and G) and structural parameters (n, B/T and $\Sigma_{1}$) show the monotonic decreasing or increasing tendency from blue to green 1 to green 2 to red galaxies, suggesting that the buildup of bulge component play an important role during the morphological transformation.

(4) For the galaxies with the same color, massive galaxies have more prominent bulge property. The quenching rate is higher in high environmental richness than that of low environmental richness if we consider the percentage of red galaxies as a indicator for quenching rate in different environments.

(5) Combining additional blue and red galaxies, we find there is a strong negative (positive) relationship between B/T and sSFR ($D_{4000}$). In the same range of $u-r$ color, massive galaxies have higher B/T when given the sSFR or $D_{4000}$. The monotonic variation suggests that the physical progresses of strengthening bulge are also the ones of weakening the star formation.

\begin{acknowledgements}
This work is supported by the National Key Research and Development Program of China (No. 2017YFA0402703) and by the National Natural Science Foundation of China (No. 11733002).
Funding for the Sloan Digital Sky Survey (SDSS)
has been provided by the Alfred P. Sloan Foundation, the Participating
Institutions, the National Science Foundation, the US
Department of Energy, the National Aeronautics and Space Administration,
the Japanese Monbukagakusho, the Max Planck
Society.
The SDSS Web site is http://www.sdss.org/. The SDSS is managed
by the Astrophysical Research Consortium for the Participating
Institutions. The Participating Institutions are the American
Museum of Natural History, Astrophysical Institute Potsdam,
University of Basel, Cambridge University, Case Western Reserve
University, University of Chicago, Drexel University, Fermilab,
the Institute for Advanced Study, the Japan Participation
Group, Johns Hopkins University (JHU), the Joint Institute for Nuclear
Astrophysics, the Kavli Institute for Particle Astrophysics and
Cosmology, the Korean Scientist Group, the Chinese Academy of
Sciences (LAMOST), Los Alamos National Laboratory, the Max-Planck-Institute for Astronomy (MPIA), the Max-Planck-Institute
for Astrophysics (MPA), NewMexico State University, Ohio State
University, University of Pittsburgh, University of Portsmouth,
Princeton University, the United States Naval Observatory, and the
University of Washington.
\end{acknowledgements}

%\appendix                  %%appendicial material is supported

\label{lastpage}


\begin{thebibliography}{99}
%% you can type \apj for ApJ, \aap for A&A, \apss for Ap&SS, etc. Please consult
%% the macro chjaa.cls. You can also find them in aasguide.tex (AASTeX for ApJ, AJ, PASP)
%% Please follow the format of ChJAA's reference list
\bibitem[Abaajian et al.(2009)]{Abazajian2009} Abazajian K. N. et al., 2009, ApJS, 182, 543

\bibitem[Abraham et al.(1996)]{Abraham1996} Abraham R. G., Tanvir N. R. et al., 1996, MNRAS, 279, L47

\bibitem[Bait et al.(2017)]{Bait2017} Bait O., Barway S., Wadadekar Y. 2017, MNRAS, 471, 2687

\bibitem[Balogh et al.(1999)]{Balogh1999} Balogh M. L. et al., 1999, ApJ, 527, 54

\bibitem[Balogh et al.(2011)]{Balogh2011} Balogh M. L. et al., 2011, MNRAS, 412, 2303

\bibitem[Bamford et al.(2009)]{Bamford2009} Bamford S. P., Nichol R. C., Baldry I. K. et al., 2009, MNRAS, 393, 1324

\bibitem[Bell et al.(2004)]{Bell2004} Bell E. F. et al., 2004, ApJ, 608, 752

\bibitem[Bell(2008)]{Bell2008} Bell E. F. 2008, ApJ, 682, 355

\bibitem[Bell et al.(2012)]{Bell2012} Bell E. F. et al., 2012, ApJ, 753, 167

\bibitem[Blanton et al.(2003)]{Blanton2003} Blanton M. R. et al., 2003, ApJ, 594, 186

\bibitem[Blanton et al.(2005)]{Blanton2005} Blanton M. R., Schlegel D. J., Strauss M. A. et al., 2005, AJ, 129, 2562

\bibitem[Bremer et al.(2018)]{Bremer2018} Bremer M. N. et al., 2018, MNRAS, 476, 12

\bibitem[Brennan et al.(2017)]{Brennan2017} Brennan R. et al., 2017, MNRAS, 465, 619

\bibitem[Brinchmann et al.(2004)]{Brinchmann2004} Brinchmann J., Charlot S., White S. D. M. et al., 2004, MNRAS, 351, 1151

\bibitem[Cameron(2011)]{Cameron2011} Cameron E. 2011, PASA, 28, 128

\bibitem[Cheung et al.(2012)]{Cheung2012} Cheung E. et al., 2012, ApJ, 760, 131

\bibitem[Chilingarian et al.(2012)]{Chilingarian2012} Chilingarian I. V. et al., 2012, MNRAS, 405, 1049

\bibitem[Coenda et al.(2018)]{Coenda2018} Coenda V. et al., 2018, MNRAS, 473, 5617

\bibitem[Cowie et al.(1996)]{Cowie1996} Cowie L. L. et al., 1996, AJ, 112, 839

\bibitem[Cutri et al.(2014)]{Cutri2014} Cutri R. M. et al., 2014, yCat, 2328, 0

\bibitem[Di Matteo et al.(2005)]{Di Matteo2005} Di Matteo T., Springel V., Hernquist L. 2005, Nature, 433, 604

\bibitem[Driver et al.(2006)]{Driver2006} Driver S. P. et al., 2006, MNRAS, 368, 414

\bibitem[Elbaz et al.(2007)]{Elbaz2007} Elbaz D. et al., 2007, A\&A, 468, 33

\bibitem[Faber et al.(2007)]{Faber2007} Faber S. M., Willmer C. N. A., Wolf C. et al., 2007, ApJ, 665, 265

\bibitem[Fang et al.(2013)]{Fang2013} Fang J. J., Faber S. M., Koo D. C., Dekel A. 2013, ApJ, 776, 63

\bibitem[Hopkins et al.(2010)]{Hopkins2010} Hopkins P. F., Bundy K., Croton D. et al., 2010, ApJ, 715, 202

\bibitem[Jin et al.(2014)]{Jin2014} Jin S. W., Gu Q. S. et al., 2014, ApJ, 787, 63

\bibitem[Kauffmann et al.(2003a)]{Kauffmann2003a} Kauffmann G. et al., 2003a, MNRAS, 341, 33

\bibitem[Kauffmann et al.(2003b)]{Kauffmann2003b} Kauffmann G. et al., 2003b, MNRAS, 341, 54

\bibitem[Kaviraj(2014a)]{Kaviraj2014a} Kaviraj S. 2014a, MNRAS, 437, L41

\bibitem[Kaviraj(2014b)]{Kaviraj2014b} Kaviraj S. 2014b, MNRAS, 440, 2944

\bibitem[Kong et al.(2009)]{Kong2009} Kong X., Fang G., Arimoto N., \& Wang M. 2009, ApJ, 702, 1458

\bibitem[Lackner \& Gunn(2012)]{Lackner2012} Lackner C. N., Gunn J. E. 2012, MNRAS, 421, 2277

\bibitem[Lang et al.(2014)]{Lang2014} Lang P., Wuyts S. et al., 2014, ApJ, 788, 11

\bibitem[Lintott et al.(2011)]{Lintott2011} Lintott C. et al., 2011, MNRAS, 410, 166

\bibitem[Lotz et al.(2004)]{Lotz2004} Lotz J. M., Primack J., \& Madau P. 2004, AJ, 128, 163

\bibitem[Marasco et al.(2012)]{Marasco2012} Marasco A., Fraternali F., Binney J. J. 2012, MNRAS, 419, 1107

\bibitem[Martin et al.(2005)]{Martin2005} Martin D. C., Fanson J., Schiminovich D. et al., 2005, ApJL, 619, L1

\bibitem[Martig et al.(2009)]{Martig2009} Martig M., Bournaud F., Teyssier R. et al., 2009, ApJ, 707, 250

\bibitem[Martig et al.(2013)]{Martig2013} Martig M., Crocker A. F., Bournaud F. et al., 2013, MNRAS, 432, 1914

\bibitem[Meert et al.(2015)]{Meert2015} Meert A., Vikram V., Bernardi M. 2015, MNRAS, 446, 3943

\bibitem[Meert et al.(2016)]{Meert2016} Meert A., Vikram V., Bernardi M. 2016, MNRAS, 455, 2440

\bibitem[Mendez et al.(2011)]{Mendez2011} Mendez A. J. et al., 2011, ApJ, 736, 110

\bibitem[Nandra et al.(2007)]{Nandra2007} Nandra K. et al., 2007, ApJ, 660, L11

\bibitem[Pan et al.(2013)]{Pan2013} Pan Z. Z. et al., 2013, ApJ, 776, 14

\bibitem[Peng et al.(2002)]{Peng2002} Peng C. Y., Ho L. C., Impey C. D. et al., 2002, AJ, 124, 266

\bibitem[Peng et al.(2010)]{Peng2010} Peng Y. J., Lilly S. J., Kova\u{c} K. et al., 2010, ApJ, 721, 193

\bibitem[Salim et al.(2007)]{Salim2007} Salim S. et al., 2007, ApJS, 173, 267

\bibitem[Salim et al.(2009)]{Salim2009} Salim S. et al., 2009, ApJ, 700, 161

\bibitem[Schiminovich et al.(2007)]{Schiminovich2007} Schiminovich D., Wyder T. K., Martin D. C. et al., 2007, ApJS, 173, 315

\bibitem[Simard et al.(2011)]{Simard2011} Simard L., Mendel J. T., Patton D. R., Ellison S. L., McConnachie A. W. 2011, ApJS, 196, 11

\bibitem[Springel et al.(2005)]{Springel2005} Springel V., Di Matteo T., Hernquist L. 2005, ApJ, 620, L79

\bibitem[Strateva et al.(2001)]{Strateva2001} Strateva I. et al., 2001, AJ, 122, 1861

\bibitem[van den Bosch(2002)]{van den Bosch2002} van den Bosch F. C. 2002, MNRAS, 331, 98

\bibitem[Vikram et al.(2010)]{Vikram2010} Vikram V. et al., 2010, MNRAS, 409, 1379

\bibitem[Wyder et al.(2007)]{Wyder2007} Wyder T. K., Martin D. C., Schiminovich D. et al., 2007, ApJS, 173, 293

\bibitem[Yang et al.(2007)]{Yang2007} Yang X., Mo H. J. et al., 2007, ApJ, 671, 153

\bibitem[Yesuf et al.(2014)]{Yesul2014} Yesuf H. M. et al., 2014, ApJ, 792, 84

\bibitem[York et al.(2000)]{York2000} York D. G. et al., 2000, AJ, 120, 1579

\end{thebibliography}
\end{document}